\documentclass[aps,prl,10pt,twocolumn,superscriptaddress]{revtex4-1}

\usepackage{times}
\usepackage{changes}
\usepackage{graphicx}
\usepackage{amssymb}
\usepackage{epstopdf}
\usepackage{amsmath}
\usepackage{color}
\usepackage{hyperref}
\usepackage{bbold}
\usepackage{enumitem}
\usepackage{bm}
\usepackage[acronym]{glossaries}
\usepackage{physics}
\makeatletter
\graphicspath{{Images/}}

\usepackage{hyperref}
\hypersetup{
	colorlinks=true,
	linkcolor=black,          % color of internal links
	citecolor=blue,           % color of links to bibliography
	filecolor=magenta,        % color of file links
	urlcolor=cyan
}

\makeatother

\begin{document}
	
\title{Classical Prethermal Phases of Matter}

\author{Andrea Pizzi}
\affiliation{Cavendish Laboratory, University of Cambridge, Cambridge CB3 0HE, United Kingdom}
\author{Andreas Nunnenkamp}
\affiliation{School of Physics and Astronomy and Centre for the Mathematics and Theoretical Physics of Quantum Non-Equilibrium Systems, University of Nottingham, Nottingham, NG7 2RD, United Kingdom}
\author{Johannes Knolle}
\affiliation{Department of Physics, Technische Universit{\"a}t M{\"u}nchen, James-Franck-Stra{\ss}e 1, D-85748 Garching, Germany}
\affiliation{Munich Center for Quantum Science and Technology (MCQST), 80799 Munich, Germany}
\affiliation{Blackett Laboratory, Imperial College London, London SW7 2AZ, United Kingdom}

\begin{abstract}
	Systems subject to a high-frequency drive can spend an exponentially long time in a prethermal regime, in which novel phases of matter with no equilibrium counterpart can be realized. Due to the notorious computational challenges of quantum many-body systems, numerical investigations in this direction have remained limited to one spatial dimension, in which long-range interactions have been proven a necessity. Here, we show that prethermal non-equilibrium phases of matter are not restricted to the quantum domain. Studying the Hamiltonian dynamics of a large three-dimensional lattice of classical spins, we provide the first numerical proof of prethermal phases of matter in a system with short-range interactions. Concretely, we find higher-order as well as fractional discrete time crystals breaking the time-translational symmetry of the drive with unexpectedly large integer as well as fractional periods. Our work paves the way towards the exploration of novel prethermal phenomena by means of classical Hamiltonian dynamics with virtually no limitations on the system's geometry or size, and thus with direct implications for experiments.
\end{abstract}

\maketitle

\textit{Introduction.---}
In the past few years, a great deal of attention has been devoted to the realization of novel phases of matter away from thermal equilibrium. The most prominent example is that of discrete time crystals (DTCs), systems that break the discrete time-translational symmetry of a periodic drive by showing a robust subharmonic response \cite{sacha2015modeling, khemani2016phase, else2016floquet, yao2017discrete}. A major impediment in the quest for nontrivial non-equilibrium phases of matter has been the fact that generic many-body systems under a periodic drive tend to quickly heat up to a featureless infinite-temperature state. Established loopholes to evade this fate are many-body localization (MBL) \cite{khemani2016phase, else2016floquet, yao2017discrete}, infinite-range interactions \cite{sacha2015modeling, russomanno2017floquet}, and dissipation \cite{gong2018discrete,lazarides2020time}.

An alternative mechanism to prevent heating has been more recently put forward: prethermalization \cite{berges2004prethermalization, mori2016rigorous, abanin2017effective, else2017prethermal,machado2020long, luitz2020prethermalization, zhao2021random}. According to this phenomenon, when the frequency $\omega = 2\pi/T$ of the drive is large, the system remains stuck in a prethermal regime for an exponentially long time $\sim e^{c\omega}$ ($c$ being some constant), before ultimately meeting its heat death \cite{canovi2016stroboscopic, abanin2017effective, weidinger2017floquet, abanin2017rigorous, mallayya2019prethermalization}. In contrast to MBL, prethermalization requires no disorder and occurs in any dimensionality, features that make it an excellent candidate for experimental implementation. The only price to pay is a finite lifetime, which for essentially all current implementations can nonetheless be tuned orders of magnitude larger than the achievable coherence times.

\begin{figure}[!bth]
	\begin{center}
		\includegraphics[width=\linewidth]{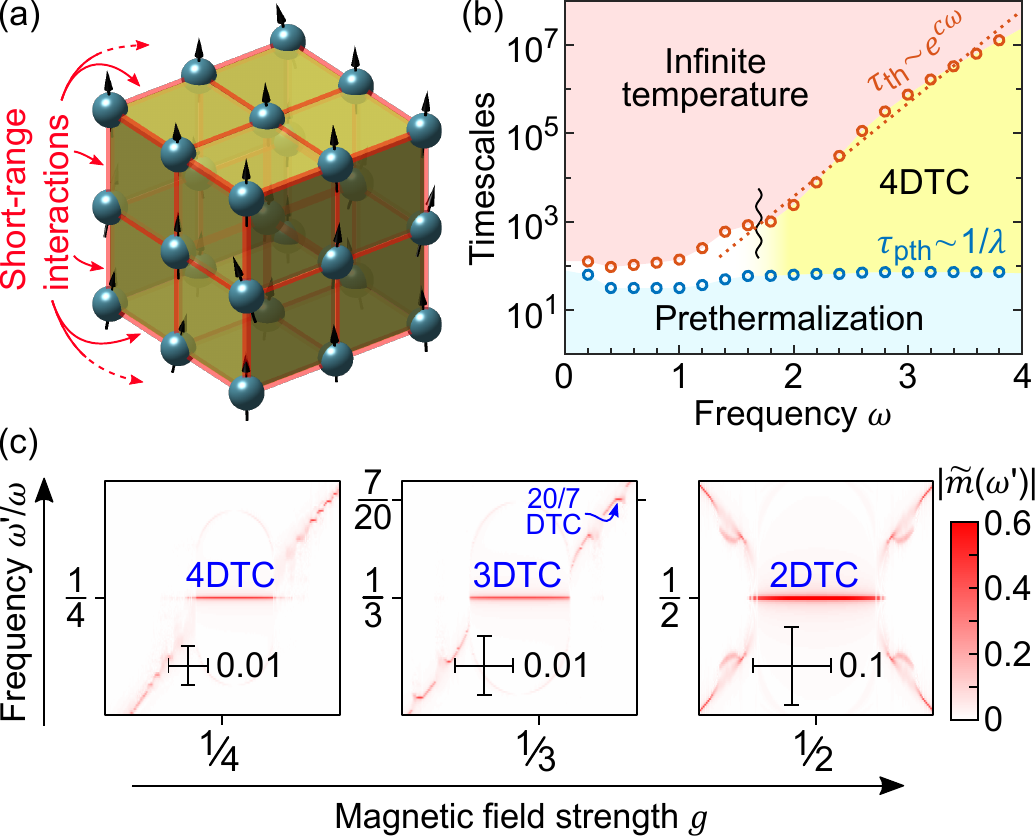}\\
	\end{center}
	\vskip -0.5cm \protect
	\caption{
		\textbf{Prethermal discrete time crystals}.
		(\textbf{a}) The system consists of $N = 50^3$ classical spins in a three-dimensional lattice with nearest-neighbor interactions. (\textbf{b}) Relevant timescales of a $4$-DTC (for $g = 0.25$). The prethermalization timescale $\tau_{pth} \sim 1/\lambda$ is set by the Lyapunov exponent $\lambda$. Thermalization to an infinite-temperature state occurs at a much later time $\tau_{th} \sim e^{c\omega}$ exponential in the drive frequency $\omega$. Here, $\Delta = 0.01$.
		(\textbf{c}) By varying the parameter $g$, one can access different higher-order and fractional $n$-DTCs, each corresponding to a plateau at frequency $1/n$ in the system's spectral response. This is shown by means of the Fourier transform $|\tilde{m}|$ of the magnetization for $n = 2,3,4$ and $20/7$.}
	\label{fig1}
\end{figure}

Else \emph{et al.}~\cite{else2017prethermal} have shown that such a prethermal regime can be exploited to realize nontrivial out-of-equilibrium phases of matter. The analytical work of Ref.~\cite{else2017prethermal} assumes short-range interactions, for which phenomena like prethermal DTCs require a dimensionality two or three. This in turn makes the important task of numerically validating the theory, its assumptions, and limitations, extremely difficult. Indeed, this has only been possible for a recent generalization to long-range one-dimensional systems \cite{machado2020long} and for relatively small system sizes.

Needless to say, working with small system sizes and in one dimension represent a major setback for both the characterization of known collective dynamical phenomena and the exploration of novel ones. As a striking example, in one dimension the signatures of higher-order and fractional DTCs only appear at system sizes exceeding by a factor 2 those in the reach of exact diagonalization (ED) techniques \cite{pizzi2021higher}. For a system of spin $1/2$, unlike standard period doubling DTCs, higher-order and fractional DTCs are characterized by a robust subharmonic response at frequency $\omega/n$ with integer and possibly even fractional $n>2$. Recently, we have characterized these exotic prethermal non-equilibrium phases of matter in a clean (that is, non-disordered) long-range one-dimensional quantum spin system \cite{pizzi2021higher}.

Interestingly, some recent studies have shown that the phenomenon of prethermalization is not unique to quantum systems \cite{rajak2018stability, mori2018floquet, rajak2019characterizations, howell2019asymptotic}, and that the concept of prethermal Hamiltonian can be extended to the classical setting \cite{mori2018floquet, howell2019asymptotic}. This suggests that the picture for prethermal DTCs drawn in Ref.~\cite{else2017prethermal} should generalize to classical Hamiltonian dynamics, which would tear down the stringent numerical constraint mentioned above, and open the way to large-scale simulations of these non-equilibrium collective phenomena.

Here, we show that this is indeed the case. We consider a clean three-dimensional system of classical spins and show that it can host prethermal higher-order and fractional DTCs for short-range (nearest-neighbor) interactions, see Fig.~\ref{fig1}. The resulting Hamiltonian (thus, non-dissipative) dynamics is dominated by two timescales. The first is related to the prethermalization of the system to an effective Hamiltonian $H_\textrm{eff}$, that occurs over a timescale $\tau_{pth} \sim 1/\lambda$, with $\lambda$ the Lyapunov exponent independent of $\omega$. The second is related to the infinite-temperature thermalization, that occurs only after an exponentially long time $\tau_{th} \sim e^{c \omega}$. The separation of timescales leaves room for the realization of prethermal $n$-DTCs with various orders $n = 2, 3, 4, 20/7$ and beyond.

We note that the notion of classical DTCs has also been adopted with various connotations by previous works \cite{gambetta2019classical, khasseh2019many, heugel2019classical, yao2020classical, malz2021seasonal, pizzi2021bistability}. We also emphasize that, in contrast to the well-known instances of classical synchronization, period doubling bifurcations, and other related phenomena in dynamical system theory \cite{strogatz2018nonlinear}, the focus of our work are many-body systems undergoing driven but non-dissipative (i.e., non-contractive) dynamics and still evading (up to a prethermal regime) the fate of ergodicity.

\textit{Model.---}
We consider a simple cubic lattice with $N$ sites, in which each site $\bm{r}$ hosts a classical spin $\bm{S}_{\bm{r}}=\left(S_{\bm{r}}^x, S_{\bm{r}}^y, S_{\bm{r}}^z\right)$ with $\left| \bm{S}_{\bm{r}} \right| = 1$. A remarkably large system size $N = 50^3$ ensures results well representative of the thermodynamic limit (as further supported by a scaling analysis in the Supplementary Information). The spins are governed by the following periodic binary Hamiltonian $H(t)$ at frequency $\omega = 2\pi/T$
\begin{equation}
H(t) = 
\begin{cases}
\frac{1}{6} \sum_{\langle \bm{r}, \bm{r}' \rangle} S_{\bm{r}}^z S_{\bm{r}'}^z
+ h \sum_{\bm{r}} S_{\bm{r}}^z \quad & \text{for} \ 0 \le t < \frac{T}{2} \\
2 \omega g \sum_{\bm{r}} S_{\bm{r}}^x \quad & \text{for} \ \frac{T}{2} \le t < T.
\end{cases}
\label{eq. H}
\end{equation}
The first part of the Hamiltonian in Eq.~\eqref{eq. H} accounts for a nearest-neighbor $ZZ$ interaction together with a longitudinal field of strength $h = 0.1$, whereas the second part describes the action of a transverse field of strength $2g\omega$. The parametrization of the latter has been chosen such that the rotation around the $x$ axis caused by the transverse field is equal to $2\pi g$ irrespective of $\omega$. For instance, when $g$ is equal to $0.5$, the second part of the Hamiltonian acts as a $\pi$-flip of the spins.

\begin{figure*}[bth]
	\begin{center}
		\includegraphics[width=\linewidth]{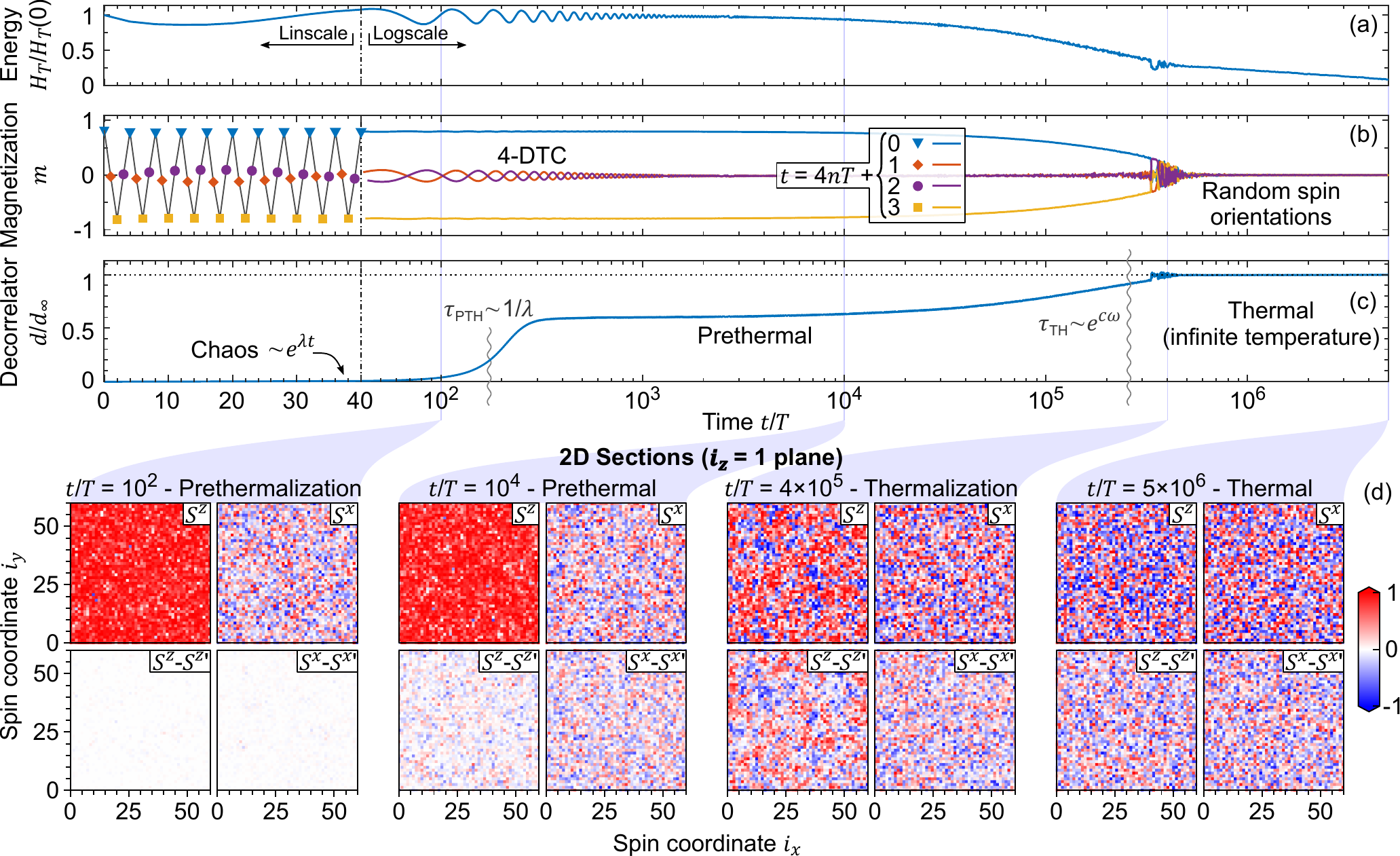}\\
	\end{center}
	\vskip -0.5cm \protect
	\caption{
		\textbf{Phenomenology of a prethermal $4$-DTC}. (\textbf{a}) The average energy $H_T$ reaches its infinite-temperature value $0$ after an exponentially long time, the signature of prethermalization. (\textbf{b}) A period $4$-tupling is observed in the stroboscopic series of the magnetization $m$ over the whole prethermal regime. (\textbf{c}) The decorrelator initially grows exponentially, $d\sim e^{\lambda t}$, signalling sensitivity to the initial conditions and chaos. After a timescale $\tau_{pth} \sim 1/\lambda$, it plateaus at a value $\sim 60\% d_\infty$, before ultimately reaching its infinite-temperature value $d_\infty = \sqrt{2}$ after a time $\tau_{th} \sim e^{c\omega}$. (\textbf{d}) Spin components $S_{\bm{r}}^{x,z}$ over a two-dimensional cut of the system and at representative times $t$ multiples of $4T$, together with the difference $S_{\bm{r}}^{x,z}-S_{\bm{r}}^{x,z\prime}$ between the two initially close copies of the system. At time $t = 10^2 T < \tau_{pth}$, the spins are predominantly polarized along $z$ ($m>0$) while having disordered $x$ components, and the difference between the two copies is still small (mostly white). At time $\tau_{pth} \ll t = 10^4T \ll \tau_{th}$ the system has prethermalized: while both copies of the system are still polarized along $z$, their $x$ and $y$ components have decorrelated. At time $t = 4 \times 10^5 T \sim \tau_{th}$, the $z$ polarization is progressively destroyed by the proliferation of domain walls, en route towards the ultimate heat death with completely random spin orientations, shown for $t = 5 \times 10^6 \gg \tau_{th}$. Here, $\omega = 2.86, g = 0.255,$ and $\Delta = 0.01$.}
	\label{fig2}
\end{figure*}

The spin dynamics is given by standard Hamilton equations of motion $\dot{S}_{\bm{r}}^{\alpha} = \left\{S_{\bm{r}}^\alpha, H(t)\right\}$, where $\left\{ \dots \right\}$ denotes Poisson brackets and $\left\{ S_{\bm{r}}^\alpha, S_{\bm{r'}}^\beta \right\} = \delta_{\bm{r}, \bm{r'}} \epsilon_{\alpha, \beta, \gamma} S_{\bm{r}}^\gamma$. As noted by Howell and collaborators for the analogue one-dimensional case \cite{howell2019asymptotic}, the resulting $3N$ coupled, nonlinear, ordinary differential equations can be integrated analytically over the two halves of the drive. Indeed, one finds
\begin{equation}
\bm{S}_{\bm{r}}(nT+T) = 
\begin{pmatrix}
1 & 0 & 0 \\
0 & c_2 & -s_2 \\
0 & s_2 & c_2 \\
\end{pmatrix}
\begin{pmatrix}
c_1 & -s_1 & 0 \\
s_1 & c_1 & 0 \\
0 & 0 & 1 \\
\end{pmatrix}
\bm{S}_{\bm{r}}(nT)
\label{eq. map}
\end{equation}
with $c_1 = \cos \left( \kappa_{\bm{r}} T/2\right)$, $s_1 = \sin \left( \kappa_{\bm{r}} T/2\right)$, $c_2 = \cos 2 \pi g$, and $s_2 = \sin 2 \pi g$. The system's ``many-bodyness'' is imprinted in the non-linearity of the equations, now hidden in the effective field $\kappa_{\bm{r}} = h + \frac{1}{6}\sum_{\bm{r}' \in \partial_{\bm{r}}} S_{\bm{r}'}^z$, with the sum running over the $6$ nearest neighbors of site $\bm{r}$. By iteratively applying the discrete map in Eq.~\eqref{eq. map} we can evolve the system up to remarkably large times $\sim 10^7 T$.

As initial condition, we consider one in which the spins are predominantly polarized along the $z$ direction. In spherical coordinates $\bm{S}_{\bm{r}} = \left( \sin \theta_{\bm{r}} \cos \phi_{\bm{r}}, \sin \theta_{\bm{r}} \sin \phi_{\bm{r}}, \cos \theta_{\bm{r}} \right)$, for every spin $\bm{S}_{\bm{r}}(0)$ the initial polar and azimuthal angles $\theta_{\bm{r}}(0)$ and $\phi_{\bm{r}}(0)$ are drawn at random from a Gaussian distribution with mean $0$ and standard deviation $2\pi W$ and from a uniform distribution between $0$ and $2\pi$, respectively. For $W = 0$ the spins are perfectly aligned along $z$ and, because of translational invariance, behave all in the same way, reducing the system to an effective single-body one. The many-body character of the system is brought into play scrambling the initial condition with a finite $W$, that can be thought of as a sort of initial `temperature'. Henceforth, $W=0.1$.

The main observables of interest are the average (over one period) energy
$H_T = \frac{1}{12} \sum_{\langle \bm{r} \bm{r}' \rangle} S_{\bm{r}}^z S_{\bm{r}'}^z + \sum_{\bm{r}} \left( \frac{h}{2} S_{\bm{r}}^z + \omega g S_{\bm{r}}^x \right)$, the magnetization $m = \frac{1}{N}\sum_{\bm{r}} S_{\bm{r}}^z$, and its Fourier transform $\tilde{m}(\omega') = \frac{1}{M} \sum_{n = 0}^{M-1} m(nT)e^{-i \omega' nT}$. Furthermore, we probe the hallmark of chaos, sensitivity to the initial conditions, by introducing a `decorrelator' $d$ that measures the distance between two initially very close copies of the system \cite{bilitewski2018temperature, bilitewski2020classical}. We define
\begin{equation}
d(t) = \sqrt{\frac{1}{N} \sum_{\bm{r}} \left(\bm{S}_{\bm{r}}(t) - \bm{S}_{\bm{r}}'(t) \right)^2},
\end{equation}
where the primed spins refer to a copy of the system that has been initially slightly perturbed. We consider $\theta_{\bm{r}}'(0) = \theta_{\bm{r}}(0) + 2\pi\Delta\delta_{\theta,\bm{r}}$ and $\phi_{\bm{r}}'(0) = \phi_{\bm{r}}(0) + 2\pi\Delta\delta_{\phi,\bm{r}}$, with $\delta_{\theta,\bm{r}}$ and $\delta_{\phi,\bm{r}}$ standard normal random numbers and $\Delta \ll 1$ setting the size of the initial perturbation [thus, $d(0)$]. At infinite temperature, when the spin orientations are completely random, the decorrelator takes a value $d_\infty = \sqrt{2}$ (see Supplementary Information).

\begin{figure}[bth]
	\begin{center}
		\includegraphics[width=\linewidth]{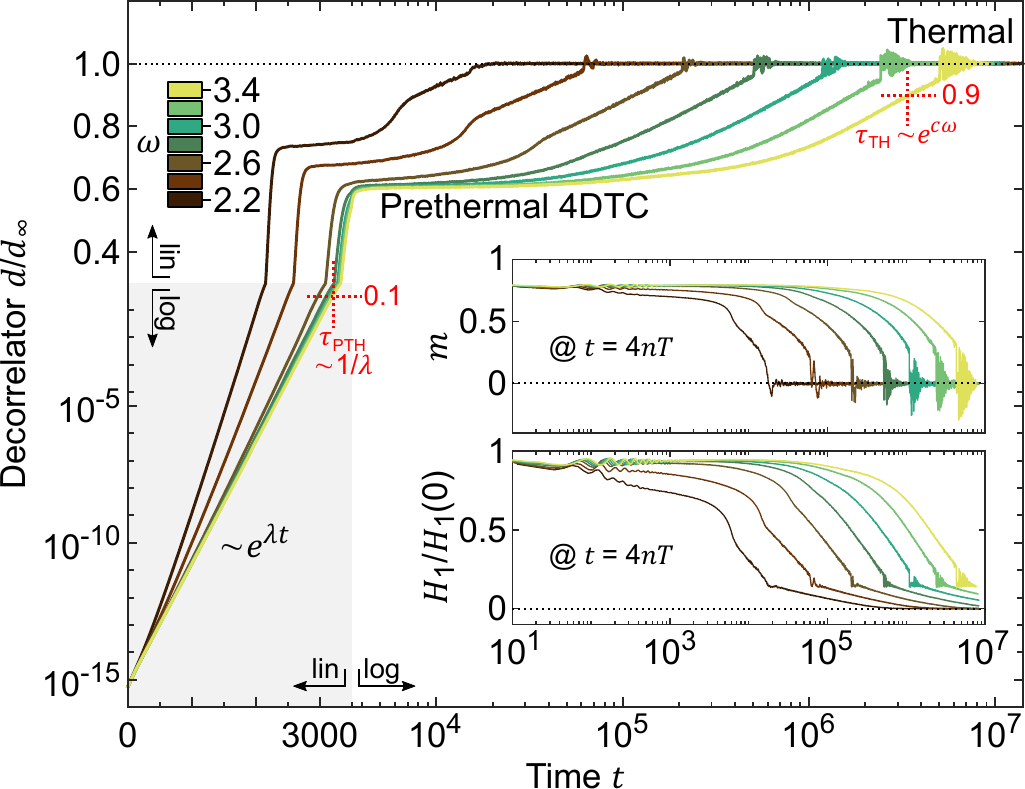}\\
	\end{center}
	\vskip -0.5cm \protect
	\caption{
		\textbf{Frequency dependence of a prethermal discrete time crystal}. The decorrelator initially grows exponentially, $d \sim e^{\lambda t}$. The prethermalization time $\tau_{pth} \sim 1/\lambda$ is associated to the time at which $d$ crosses the $10\%$ of its infinite-temperature value $d_\infty$, and weakly depends on the frequency $\omega$. By contrast, the prethermal plateau $d \approx 60\% d_\infty$ extends for a time that grows exponentially with the frequency. As a reference for the thermalization time $\tau_{th}$, we take the time at which the decorrelator $d$ crosses $90\%d_\infty$. Note, the axes have linear and logarithmic parts. The insets show the magnetization $m$ and the energy $H_1$ [the first part of $H(t)$ in Eq.~\ref{eq. H}] at stroboscopic times $t = 4nT$. Here, $g = 0.25$ and $\Delta = 10^{-16}$.}
	\label{fig3}
\end{figure}

\textit{Results.---} We start by shading light on the zoology of possible DTCs. To this end, in Fig.~\ref{fig1}(c) we plot the magnetization Fourier transform $\tilde{m}(\omega')$ as a function of $\omega'$ and $g$. A constant-frequency plateau signals a robust DTC. The plateau frequency $\omega/n$ indicates the order $n$ of the DTC (here, we illustrate the prime cases $n = 2,3,4$ and $20/7$), whereas its width signals its stability to perturbations of $g$ \cite{pizzi2021higher}. In the remainder of the paper, we focus for concreteness on the properties of the $4$-DTC, obtained for $g \approx 1/4$.

To elucidate the phenomenology of prethermalization and time crystallinity, in Fig.~\ref{fig2} we show the time series of the observables introduced above. First and foremost, prethermalization is diagnosed by looking at the average energy in Fig.~\ref{fig2}(a): $H_T$ plateaus at a value $\sim H_T(0)$ over $\sim 10^5$ decades, before heating ultimately takes it to its infinite-temperature value $0$. Crucially, prethermalization comes along with the realization of a nontrivial non-equilibrium phase of matter, the $4$-DTC, that can be diagnosed by looking at the magnetization $m$ in Fig.~\ref{fig2}(b). At short times $t/T < 40$, a linear axis helps to appreciate the distinctive stroboscopic dynamics of a $4$-DTC: the magnetization takes values $\sim 1,0,-1,0,1,0,\dots$ at $t/T = 0,1,2,3,4,5,\dots$, thus exhibiting a characteristic frequency $\omega/4$. For $t/T > 40$, the logarithmic time axis allows to assess the persistence of the subharmonic response over the whole prethermal regime, before it reaches its infinite-temperature value $0$. 

The nature of the prethermal $4$-DTC is perhaps even more strikingly highlighted by the decorrelator $d$ in Fig.~\ref{fig2}(c). At short times, the decorrelator grows exponentially as $d \sim e^{\lambda t}$ according to a characteristic Lyapunov exponent $\lambda$. This sensitivity to initial conditions is the signature of chaos, the one-to-one classical correspondent of quantum thermalization. Rather than directly approaching the infinite-temperature value $d_\infty$, however, the decorrelator plateaus at a value $\sim 60 \% d_\infty$ for the whole prethermal regime.

A deeper understanding of the prethermal $4$-DTC is achieved by looking at the space profiles of the spins at representative times. To avoid complicated three-dimensional plots (see Supplementary Information), in Fig.~\ref{fig2}(d) we restrict for clarity to the plane of spins with lattice indices $(i_x, i_y, 1)$. We plot the $S^x$ and $S^z$ components of the spins, together with the respective difference $S^x-{S^x}'$ and $S^z-{S^z}'$ between the two copies of the system used to compute the decorrelator $d$. 
We identify four main regimes in the system's evolution, and consider one representative time (multiple of $4T$) for each:
(i) Prethermalization -- At short times $t/T = 10^2 \ll \tau_{pth}$, the system is thermalizing towards the prethermal state. Following from the initial condition, $S^z \sim 1$ while $S^x \sim \pm 2\pi W$. The two copies are still close ($d \ll 1, \left|S^x-{S^x}'\right|\ll 2\pi W, $ and $\left|S^z-{S^z}'\right|\ll 1$).
(ii) Prethermal -- At intermediate times $ \tau_{pth} \ll t/T = 10^4 \ll  \tau_{th}$ we observe a prethermal $4$-DTC. The spins are still polarized along $z$, $S^z \sim 1$ and $S^x \sim \pm 2\pi W$, but chaos has decorrelated the two initially close copies of the system ($d \sim 60 \% d_\infty$, $\left|S^x-{S^x}'\right| \sim 2 \pi W$).
(iii) Thermalization -- At long times $t/T = 4 \times 10^5 \sim \tau_{th}$ the $4$-DTC is melting: the spin $z$ polarization is progressively lost with the nucleation and proliferation of domains with opposite magnetization.
(iv) Thermal -- At very long times $t/T = 5 \times 10^6 > \tau_{th}$ the system has reached (or is about to reach) its infinite-temperature state with completely random spin orientations and $d \approx d_\infty$.

A closer look at the frequency dependence of the prethermal $4$-DTC is taken in Fig.~\ref{fig3}. With a perturbation $\Delta = 10^{-16}$ saturating machine precision we emphasize the exponential growth $d \sim e^{\lambda t}$ at short times. Crucially, the Lyapunov exponent $\lambda$, that quantifies the chaoticness of the system, and therefore the timescale of the prethermalization, only weakly depends on the considered frequencies (almost no dependence is observed for large enough frequencies). In striking contrast, the full thermalization timescale at which $d$ crosses over to $d_\infty$ scales exponentially with frequency. To be quantitative, we identify the prethermalization and thermalization timescales $\tau_{pth}$ and $\tau_{th}$ with the times at which $d$ crosses $10\%$ and $90\%$ of its infinite-temperature value (marked in red for $\omega = 3.4$). The different frequency dependence of $\tau_{pth}$ and $\tau_{th}$ opens up a long prethermal window, within which the $4$-DTC is stable, see Fig.~\ref{fig1} (b). In the insets we show that analogous scalings are observed for the average energy $H_T$ and magnetization $m$ measured at stroboscopic times $t = 4kT$.

\textit{Discussion and conclusions.---}
We investigated prethermal phases of matter in a clean system of classical spins on a cubic lattice with short-range interactions subject to a periodic drive. Under suitable conditions, a separation of timescales occurs such that nontrivial prethermal phases of matter emerge, which we illustrated with a whole range of higher-order and fractional DTCs. Chaos makes the system prethermalize over a timescale $\tau_{pth} \sim \frac{1}{\lambda}$, with $\lambda$ the Lyapunov exponent that we expect to be associated to a frequency independent effective Hamiltonian $H_{eff}$. The latter could be found in a theory of classical prethermalization \cite{mori2018floquet} extended to prethermal phases of matter \cite{else2017prethermal}. The time $\tau_{th}$ for the system to then reach the infinite-temperature state is exponential (or at least \emph{nearly} exponential \cite{else2017prethermal}) in frequency.

In essence, much of the physical intuition for prethermal DTCs relies on two points \cite{else2017prethermal}: (i) energy absorption is slow because of the mismatch between driving frequency and local energy scales and (ii) the effective prethermal Hamiltonian has a finite-temperature phase transition higher than that of the initial condition. This intuition does not rely on any quantum interference effect (as, by contrast, MBL DTCs do instead \cite{khemani2016phase, else2016floquet, yao2017discrete}), and is in this sense purely classical. And indeed, it can be shown that a long-range one-dimensional version of our classical spin model \cite{pizzi2021classicalb} essentially reproduces all the main features of the prethermal DTCs in the corresponding quantum models of Refs.~\cite{machado2020long} and \cite{pizzi2021higher}. All this makes us conjecture that prethermal phases of matter of DTC type can be understood as being robust to quantum fluctuations rather than dependent on them. This perspective elevates classical Hamiltonian many-body systems to a privileged position for the investigation of novel phenomena in the non-equilibrium domain, and motivates the use of the brackets around the word `classical' in the title of this paper.

Indeed, numerical simulations become incomparably more accessible in the absence of quantum fluctuations, and the trick used in Eq.~\ref{eq. map} of integrating the dynamics over each period makes them even more efficient \cite{howell2019asymptotic}. The constraints on dimensionality and system size are in this way lifted, which opens the possibility to simulate experimentally relevant settings, beyond the few numerical one-dimensional examples considering power-law interactions $\sim 1/r^\alpha$ with rather low exponents $\alpha \lessapprox 1.5$ \cite{machado2020long, pizzi2021higher}. In particular, by providing the first simulation of higher-order DTCs in short-range interacting systems, we have provided evidence that these exotic non-equilibrium phases of matter might be much easier to realize in experiments than expected. We emphasize that no disorder is needed to guarantee a stable prethermal regime, and that a high-frequency drive should suffice. We therefore expect that the rich phenomenology we discussed here might be readily observable in state-of-the-art experiments, for instance with nitrogen–vacancy (NV) spin impurities in diamond \cite{choi2017observation}, trapped atomic ions \cite{zhang2017observation}, or $^{31}$P nuclei in ammonium dihydrogen phosphate (ADP) \cite{rovny2018observation}. Higher-order and fractional DTCs offer the chance to overcome the period-doubling paradigm of MBL DTCs, opening the way to the realization of an array of new dynamical phenomena.

As an outlook on future research, it would be insightful to clarify the exact functional form of $\tau_{th}(\omega)$. An important question regards then the quantitative effects of quantum fluctuations on our findings. As we argued above, we expect that the main features of the prethermal DTCs will not be significantly changed by quantum fluctuations, and this could for instance be checked with a suitable spin-wave approximation in higher dimension \cite{pizzi2021higher,lerose2018chaotic}. However, finding genuine quantum prethermal phases with no classical counterparts will be a very worthwhile endeavour. Similarly important will be the exploration of novel prethermal phases beyond the paradigmatic DTCs.

\emph{Note added:} During the completion of this work, we became aware of complementary work exploring prethermal DTCs in classical spin systems \cite{ye2021classical}, appeared in the same arXiv posting.

\begin{acknowledgments}
	\textit{Acknowledgements.---}
	We thank R.~Moessner and H.~Zhao for interesting discussions. We acknowledge support from the Imperial-TUM flagship partnership. A.~P.~acknowledges support from the Royal Society and hospitality at TUM. A.~N.~holds a University Research Fellowship from the Royal Society.
\end{acknowledgments}


%Supplemental material
%\newpage
\clearpage

\setcounter{equation}{0}
\setcounter{figure}{0}
\setcounter{page}{1}
\thispagestyle{empty} %Remove the counter from the first page
\makeatletter 
\renewcommand{\thefigure}{S\arabic{figure}}
\renewcommand{\theequation}{S\arabic{equation}}
\setlength\parindent{10pt}

\onecolumngrid

\begin{center}
	{\fontsize{12}{12}\selectfont
		\textbf{Supplementary Information for\\Classical Prethermal Phases of Matter"\\[5mm]}}
	{\normalsize Andrea Pizzi, Andreas Nunnenkamp, and Johannes Knolle \\[1mm]}
\end{center}
\normalsize

These Supplementary Information are devoted to a few technical details and complimentary results. In Section I we compute the infinite-temperature value of the decorrelator, $d_\infty$. In Section II we perform a scaling analysis to investigate finite-size effects and corroborate our results by showing a three-dimensional version of Fig.~2d.

\section{I) Decorrelator at infinite temperature}
In this Section, we compute the infinite-temperature value of the decorrelator $d_\infty = \sqrt{2}$. At infinite temperature, the spins are completely uncorrelated randomly orientated, so that, in the thermodynamic limit $N \to \infty$, the sum in Eq.~3 in the main text can be interpreted as an average over the possible random orientations of the spins. Moreover, no direction is preferential, and we can therefore write
\begin{equation}
    d_\infty = \lim_{N \to \infty} \left( \sqrt{\frac{1}{N} \sum_{\bm{r}} \left(\bm{S}_{\bm{r}}(t) - \bm{S}_{\bm{r}}'(t) \right)^2} \right)_{\substack{\text{random} \\ \text{spins}}}
    \rightarrow
    \sqrt{3 \int \frac{d\Omega_1}{4\pi} \frac{d\Omega_2}{4\pi} \left(S^z(\Omega_1) - {S^z}'(\Omega_2)\right)^2},
\end{equation}
where the integration is performed over the solid angles $\Omega_1$ and $\Omega_2$ associated to the spins $\bm{S}$ and $\bm{S}'$. We carry out such an integration in a straightforward manner
\begin{align}
    \int \frac{d\Omega_1}{4\pi} \frac{d\Omega_2}{4\pi} \left(S^z(\Omega_1) - {S^z}'(\Omega_2)\right)^2
    & = \frac{1}{(4\pi)^2} \int d\phi_1 d\phi_2 d\theta_1 d\theta_2 \sin \theta_1 \sin \theta_2
    \left(\cos \theta_1 - \cos \theta_2\right)^2 \\
    & = \frac{1}{4} \int d\theta_1 d\theta_2 \sin \theta_1 \sin \theta_2
    \left(\cos^2 \theta_1 + \cos^2 \theta_2 - 2\cos \theta_1 \cos \theta_2\right) \\
    & = \frac{1}{4} \int d\theta_1 \sin \theta_1 \left(2 \cos^2 \theta_1 + \frac{2}{3} - 0 \times 2\cos \theta_1 \right) =  \frac{2}{3},
\end{align}
from which we get
\begin{equation}
    d_\infty = \sqrt{2},
\end{equation}
as we set ourselves to show.

\section{II) Complimentary results}
\subsection{Scaling analysis}
\begin{figure*}[bth]
	\begin{center}
		\includegraphics[width=\linewidth]{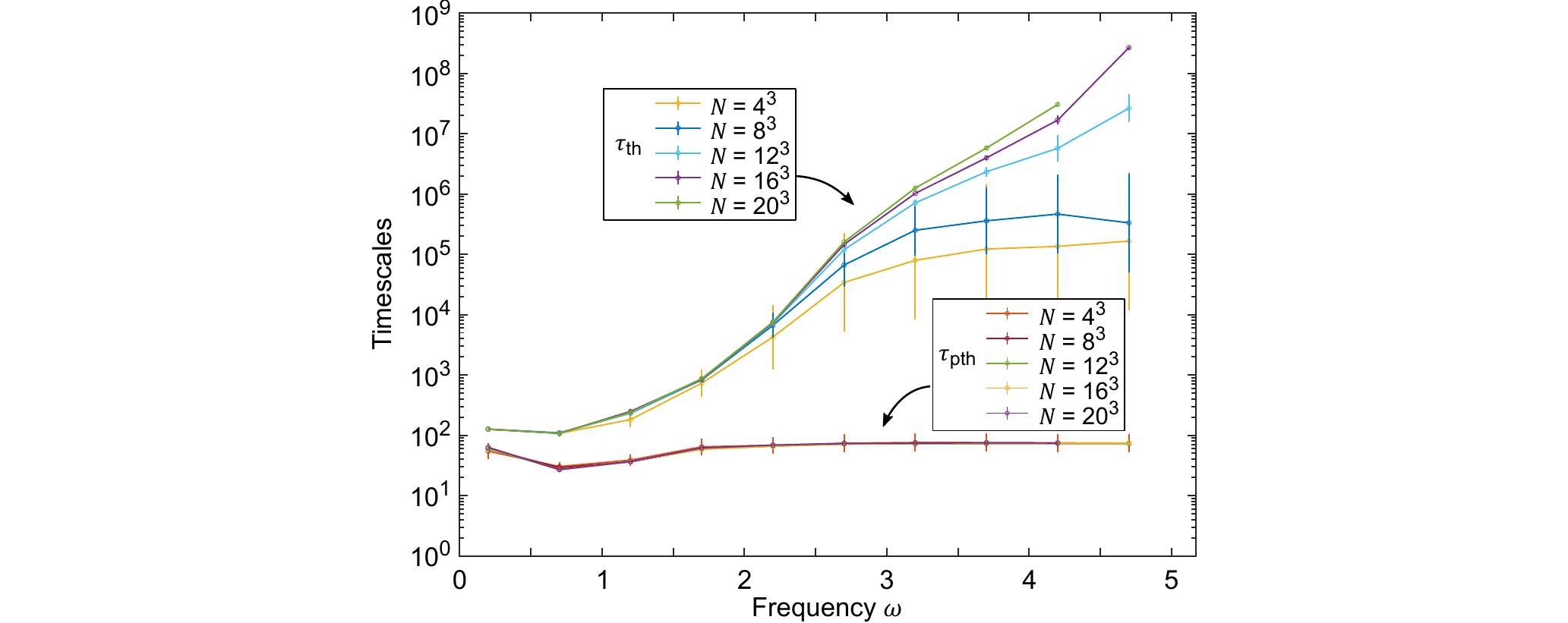}\\
	\end{center}
	\vskip -0.5cm \protect
	\caption{
		\textbf{Scaling with system size}. We plot for various system sizes $N = L^3$ the prethermalization and thermalization times of a $4$-DTC (corresponding to the $10\%$ and $90\%$ crossing times of the decorrelator, as defined in the main text). The prethermalization time $\tau_{pth}$ weakly depends on the frequency $\omega$, saturates to a constant value $\sim 1/\lambda$ for $\omega \gtrapprox 2$, and is almost independent of the system size $N$. The thermalization time $\tau_{th}$ grows with the frequency $\omega$, and depends on the system size. For small system sizes, we observe that, after an initial growth, $\tau_{th}$ saturates to a constant value. This saturation is lifted as the system size grows. In this plot, the solid lines and the error bars are associated to the average and standard deviation of the results from $R = 28^3/N$ independent runs. The parameters used are $g = 0.25, \Delta = 0.01$.}
	\label{figS1}
\end{figure*}

Here, we investigate finite size effects by looking at the timescales $\tau_{pth}$ and $\tau_{th}$ for various system sizes $N$. To this end, Fig.~\ref{figS1} reproduces Fig.~1b from the main text, but for various $N$. Since for small system sizes $N$ the statistical fluctuations are larger, to ensure a good quality of the result we consider $R \approx 28^3/N$ independent realizations of the initial conditions, and use them to compute the mean and standard deviation of $\tau_{pth}$ and $\tau_{th}$.

\subsection{Three-dimensional representations of the spin configurations}
Here, we corroborate Fig.~2d from the main text by showing the spin $x$ and $z$ components across three different planes cutting the three-dimensional volume of the system. The data and phenomenology is completely analogous to Fig.~2d, to which we refer for further details.
\begin{figure*}[bth]
	\begin{center}
		\includegraphics[width=\linewidth]{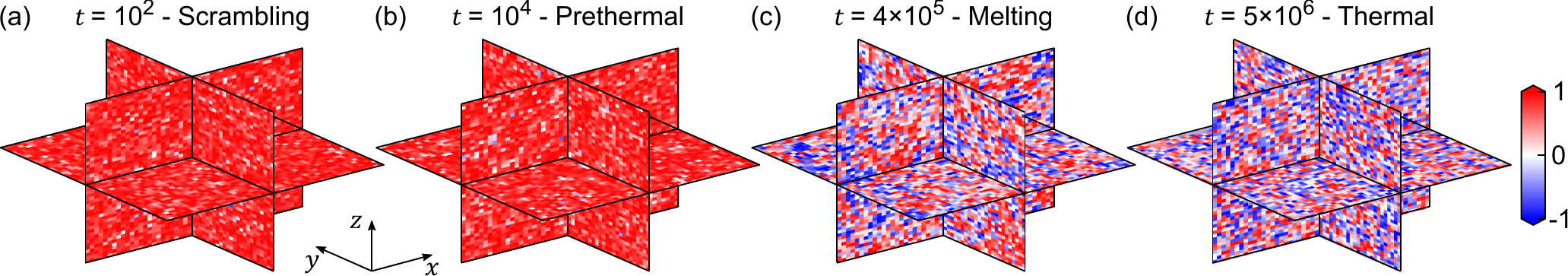}\\
	\end{center}
	\vskip -0.5cm \protect
	\caption{
		\textbf{Three-dimensional spin configurations}. Space distribution of the spins $z$ component for the same parameters considered in Fig.~2 in the main. We consider various two-dimensional cuts in the three-dimensional volume. At times $t = 10^2$ and $t = 10^4$, the spins are polarized along $z$. At time $t = 4 \times 10^5$ the spins are melting, in the sense that they exhibit a mixture of domains with different polarizations. At very long times $t = 5 \times 10^5$, the system has (almost) reached the infinite-temperature state, in which the spins' orientations are completely random.}
	\label{figS2}
\end{figure*}

\end{document}